\documentclass[preprint,showpacs,preprintnumbers,amsmath,amssymb]{revtex4}

\usepackage{graphicx}
\usepackage{dcolumn}
\usepackage{bm}

\begin{document}

\noindent

\preprint{}

\title{Understanding quantization: a hidden variable model}

\author{Agung Budiyono}
\email{agungby@yahoo.com}

\affiliation{Institute for the Physical and Chemical Research, RIKEN, 2-1 Hirosawa, Wako-shi, Saitama 351-0198, Japan}


\begin{abstract}  
We argue that to solve the foundational problems of quantum theory one has to first understand what it means to quantize a classical system. We then propose a quantization method based on replacement of deterministic c-numbers by stochastically-parameterized c-numbers. Unlike canonical quantization, the method is free from operator ordering ambiguity and the resulting quantum system has a straightforward interpretation as statistical modification of ensemble of classical trajectories. We then develop measurement without wave function collapse \`a la pilot-wave theory and point out new testable predictions.
\end{abstract}

\maketitle

\section{Introduction}

While pragmatically quantum theory has shown spectacular successes, its foundation with respect to quantum-classical correspondence and measurement problem, despite many attempts, still resist unambiguous explanation. We believe that these difficulties originate from the (physical) ambiguity of the quantization scheme through which many successful quantum systems are obtained from the corresponding classical systems. In other words, to solve the foundational problems of quantum theory, we have to first understand what is meant by quantizing a classical system. In the canonical quantization, one first writes the classical dynamical equation in Cartesian coordinate. Quantization is then done by promoting the pair of canonical conjugates variables into the corresponding Hermitian operators, $\circ\mapsto\hat{\circ}$, and Poisson bracket is replaced by the commutator: $\{\circ,\star\}\mapsto[\hat{\circ},\hat{\star}]/i\hbar$, where $\hbar$ is the reduced Planck constant. This procedure is usually said as replacement of commuting c-number (classical number) by non-commuting q-number (quantum number or Hermitian operators).  

Such direct substitution rule however implies formal, conceptual, and foundational problems. First, given a classical quantity, the above rule in general leads to infinitely many different alternatives of Hermitian operators due to operator ordering ambiguity. Second, the physical meaning of the resulting quantum systems can not be deduced ``directly'' from the quantization processes: the quantization does not tell us the physical meaning of the Schr\"odinger equation, nor it offers an explanation to the physical origin of Planck constant. In other words, the quantum-classical correspondence is not physically transparent. Hence, unlike the original classical system, we need to further introduce a physical interpretation to the resulting quantum system. Unfortunately, there are many physical interpretations with the same empirical prediction: standard Copenhagen interpretation, pilot-wave theory, many worlds etc. There is then the foundational problem that that some of the existing interpretations suffer from the infamous measurement problem \cite{Isham book}. In the present paper, we shall discuss a new method of quantization which are free from the above mentioned problems.  

\section{Quantum fluctuations as statistical modification of classical ensemble parameterized by unbiased hidden random variable}

\subsection{General formalism}

Let us consider a classical dynamics with $N$ degree of freedom whose configuration coordinate is $q=(q_1,\dots,q_N)$. Generalization to infinite degree of freedom is straightforward at least formally. In classical mechanics, all the dynamical information is contained in the Lagrangian function $\underline{L}=\underline{L}(q,\dot{q})$, where $\dot{q}\doteq dq/dt$, is the velocity. For simplicity, let us assume that the Lagrangian is non-singular. From the Hamilton principle, one then has the Euler-Lagrange equation $(d/dt)(\partial\underline{L}/\partial\dot{q}_i)-\partial\underline{L}/\partial q_i=0,\hspace{2mm}i=1,\dots,N$. To go to the Hamiltonian formalism, first, the canonical momentum is defined as $ \underline{p}_i\doteq\partial\underline{L}/\partial\dot{q}_i$, $i=1,\dots,N$, which, due to non-singularity of the Lagrangian, can be inverted to give $\dot{q}_i=\dot{q}_i(q,\underline{p};t)$, $i=1,\dots,N$. The classical Hamiltonian is defined as $\underline{H}(q,\underline{p})\doteq\underline{p}\dot{q}(q,\underline{p};t)-\underline{L}(q,\underline{p};t)$. 
The Euler-Lagrange equation can then be put into the Hamilton equation
\begin{equation}
\dot{q}_i=\frac{\partial\underline{H}}{\partial\underline{p}_i},\hspace{2mm}\dot{\underline{p}_i}=-\frac{\partial\underline{H}}{\partial q_i},\hspace{2mm} i=1,\dots,N.  
\label{Hamilton equation}
\end{equation}

Now let us consider a congruence of curves, $\{q(t;u),p(t;v)\}$, satisfying the Hamilton equation, where $u=\{u_1,\dots,u_N\}$ and $v=\{v_1,\dots,v_N\}$ are parameters of the congruences. Namely, each pair of the value of $(u,v)$ corresponds to a single member of the congruence. Let us then assume that there is a one-parameter family of hypersurfaces $\underline{S}(q;t)=\tau$, where $\underline{S}$ is a real-valued function and $\tau$ is a parameter, satisfying the following relation: 
\begin{equation}
p_i=\partial_{q_i}\underline{S}. 
\label{Hamilton-Jacobi condition}
\end{equation}
Then, it can be shown that $S(q;t)$ also satisfies the following Hamilton-Jacobi equation \cite{Rund book}:
\begin{eqnarray}
\partial_t\underline{S}+\underline{H}(q,\partial_q\underline{S})=0.
\label{Hamilton-Jacobi equation} 
\end{eqnarray}
Conversely, defining $\underline{p}$ as in Eq. (\ref{Hamilton-Jacobi condition}), the Hamilton-Jacobi equation of (\ref{Hamilton-Jacobi equation}) can be shown to lead to the Hamilton equation of (\ref{Hamilton equation}). 

Now let us consider an ensemble of copies of the system. Then, if the probability density of the position of the ensemble of trajectories in configuration space is denoted by $\underline{\rho}(q;t)$, then it must also satisfy the following continuity equation:  
\begin{equation} 
\partial_t\underline{\rho}+\partial_q\cdot(\dot{q}(\underline{S})\underline{\rho})=0, 
\label{continuity equation}
\end{equation}
where the functional form of the velocity $\dot{q}$ with respect to $\underline{S}$ is determined by substituting Eq. (\ref{Hamilton-Jacobi condition}) into the left equation of (\ref{Hamilton equation}) 
\begin{equation}
\dot{q}_i(\underline{S})=\frac{\partial\underline{H}}{\partial\underline{p}_i}\Big|_{\underline{p}=\partial_{q}\underline{S}},\hspace{2mm}i=1,\dots,N.
\label{classical velocity field} 
\end{equation}
Hence, the dynamics and statistics of the ensemble of trajectories is given by solving Eqs. (\ref{Hamilton-Jacobi equation}), (\ref{continuity equation}) and (\ref{classical velocity field}). 

Let us develop a general scheme to modify the above classical dynamics of the ensemble of trajectories \cite{AgungDQM1}. To do this, let us introduce two real-valued functions $\Omega(q,\lambda;t)$ and $S(q,\lambda;t)$ where $\lambda$ in non-vanishing hidden random variable, $\lambda\neq 0$. $\Omega(q,\lambda;t)$ is the joint-probability density of the fluctuations of $q$ and $\lambda$ so that the marginal probability densities are given by 
\begin{equation} 
\rho(q;t)\doteq\int d\lambda\Omega,\hspace{2mm}P(\lambda)\doteq\int dq\Omega,
\label{marginal probabilities}
\end{equation}
where we have assumed that the probability density of $\lambda$ is stationary. 

Now let us postulate the following rule of replacement of (deterministic) c-number by (stochastic) c-number to be applied to Eqs. (\ref{Hamilton-Jacobi equation}) and (\ref{continuity equation}) governing the dynamics of the ensemble of trajectories:
\begin{eqnarray}
\underline{\rho}\mapsto\Omega,\hspace{2mm}\partial_q\underline{S}\mapsto\partial_qS+\frac{\lambda}{2}\frac{\partial_q\Omega}{\Omega},\hspace{2mm}\partial_t\underline{S}\mapsto\partial_tS+\frac{\lambda}{2}\frac{\partial_t\Omega}{\Omega}+\frac{\lambda}{2}\partial_q\cdot\dot{q}(S),
\label{fundamental equation}
\end{eqnarray}
where the functional form of $\dot{q}(S)$ in the third line is determined by the classical Hamiltonian as in Eq. (\ref{classical velocity field}). That is we just substitute $\underline{S}$ in Eq. (\ref{classical velocity field}) with $S$. Let us show that the above rule of replacement has a consistent classical correspondence. First, expanding as $\Delta F=\partial_tF\Delta t+\partial_qF\cdot\Delta q$, using the last two equations of (\ref{fundamental equation}) one gets $\Delta\underline{S}\mapsto\Delta S+(\lambda/2)(\Delta\Omega/\Omega+\partial_q\cdot\dot{q}(S)\Delta t)$. In the limit of $S\rightarrow\underline{S}$, the second term on right hand side has to be vanishing to give $\Delta\Omega+\Omega\partial_q\cdot\dot{q}(\underline{S})\Delta t=0$. This is just the classical continuity equation so that one has $\rho=\int d\lambda\Omega\rightarrow\underline{\rho}$. 

The next question is then what is the statistics of $\lambda$. We shall show in the next subsection by taking a concrete example that the above modification of classical dynamics of ensemble of trajectories leads to the Schr\"odinger equation with unique quantum Hamiltonian if the probability density of $\lambda$ is given by  
\begin{equation}
P(\lambda)=\frac{1}{2}\delta(\lambda-\hbar)+\frac{1}{2}\delta(\lambda+\hbar). 
\label{God's coin}
\end{equation}
Namely $\lambda$ is a binary unbiased random variable which can take values $\pm\hbar$. Extension to a  more general continuous hidden random variable satisfying the unbiased condition
\begin{equation}
P(\lambda)=P(-\lambda),
\label{unbiased condition}
\end{equation}
will be given in Section 4, suggesting a very small yet finite corrections to the prediction of quantum mechanics. 

\subsection{Particle in external potentials}

Let us apply the above general formalism to a single particle subjected to external potentials. The Lagrangian is given by $\underline{L}=\frac{1}{2}g_{ij}(q)\dot{q}^i\dot{q}^j+A\cdot\dot{q}-V(q)$, where $A=(A_1,A_2,A_3)$ and $V$ are the vector and scalar potentials respectively, $g_{ij}(q)$ is an invertible matrix which might depend on $q$ (for example in the case of particle with position-dependent mass widely used in solid state physics), and summation over repeated indices is implied. Writing the inverse of $g_{ij}$ as $g^{ij}$ so that $g_{ik}g^{kl}=\delta_i^l$, the momentum is related to the velocity as 
\begin{equation}
\dot{q}^i=g^{ij}(\underline{p}_j-A_j). 
\label{classical velocity field particle in potentials}
\end{equation}
so that the classical Hamiltonian takes the following form:
\begin{equation}
\underline{H}(q,\underline{p})=\frac{g^{ij}(q)}{2}(\underline{p}_i-A_i)(\underline{p}_j-A_j)+V. 
\end{equation}

The Hamilton-Jacobi equation of (\ref{Hamilton-Jacobi equation}) then reads
\begin{equation}
\partial_t\underline{S}+\frac{g^{ij}}{2}(\partial_{q_i}\underline{S}-A_i)(\partial_{q_j}\underline{S}-A_j)+V=0.
\label{H-J equation particle in potentials}
\end{equation}
On the other hand, putting Eq. (\ref{Hamilton-Jacobi condition}) into Eq. (\ref{classical velocity field particle in potentials}), the functional relation between $\dot{q}$ and $\underline{S}$ is given by 
\begin{equation}
\dot{q}^i(\underline{S})=g^{ij}(\partial_{q_j}\underline{S}-A_j). 
\label{classical velocity field HPF particle in potentials}
\end{equation}
Inserting this into Eq. (\ref{continuity equation}), one thus has 
\begin{equation} 
\partial_t\underline{\rho}+\partial_{q_i}\Big((g^{ij}(\partial_{q_j}\underline{S}-A_j))\underline{\rho}\Big)=0. 
\label{continuity equation particle in potentials}
\end{equation}
Hence, the dynamics and statistics of the classical trajectories are determined by solving Eqs. (\ref{H-J equation particle in potentials}), (\ref{classical velocity field HPF particle in potentials}) and (\ref{continuity equation particle in potentials}). On the other hand, using Eq. (\ref{classical velocity field HPF particle in potentials}), Eq. (\ref{fundamental equation}) becomes
\begin{eqnarray}
\underline{\rho}\mapsto\Omega,\hspace{2mm}
\partial_q\underline{S}\mapsto\partial_qS+\frac{\lambda}{2}\frac{\partial_q\Omega}{\Omega},\hspace{2mm}
\partial_t\underline{S}\mapsto\partial_tS+\frac{\lambda}{2}\frac{\partial_t\Omega}{\Omega}+\frac{\lambda}{2}\partial_{q_i}g^{ij}(\partial_{q_j}S-A_j),
\label{fundamental equation particle in potentials} 
\end{eqnarray}

Now let us apply the rule of replacement of Eq. (\ref{fundamental equation particle in potentials}) to Eqs. (\ref{H-J equation particle in potentials}) and (\ref{continuity equation particle in potentials}). First, imposing the first two equations of Eq. (\ref{fundamental equation particle in potentials}) into Eq. (\ref{continuity equation particle in potentials}), one has 
\begin{equation}
\partial_t\Omega+\partial_{q_i}\Big(g^{ij}(\partial_{q_j}S-A_j)\Omega\Big)+\frac{\lambda}{2}\partial_{q_i}(g^{ij}\partial_{q_j}\Omega)=0. 
\label{FPE particle in potentials}
\end{equation}
On the other hand, inserting the last two equations of (\ref{fundamental equation particle in potentials}) into Eq. (\ref{H-J equation particle in potentials}), one has, after arrangement, 
\begin{eqnarray}
\partial_tS+\frac{g^{ij}}{2}(\partial_{q_i}S-A_i)(\partial_{q_j}S-A_j)+V-\frac{\lambda^2}{2}\Big(g^{ij}\frac{\partial_{q_i}\partial_{q_j}R}{R}+\partial_{q_i}g^{ij}\frac{\partial_{q_j}R}{R}\Big)\nonumber\\
+\frac{\lambda}{2\Omega}\Big(\partial_t\Omega+\partial_{q_i}\Big(g^{ij}(\partial_{q_j}S-A_j)\Omega \Big)+\frac{\lambda}{2m}\partial_{q_i}(g^{ij}\partial_{q_j}\Omega)\Big)=0,
\label{preHJM particle in potentials}
\end{eqnarray}
where we have defined $R\doteq\sqrt{\Omega}$ and used the following identity
\begin{equation}
\frac{1}{4}\frac{\partial_{q_i}\Omega\partial_{q_j}\Omega}{\Omega^2}=\frac{1}{2}\frac{\partial_{q_i}\partial_{q_j}\Omega}{\Omega}-\frac{\partial_{q_i}\partial_{q_j}R}{R}. 
\label{fluctuations decomposition}
\end{equation}
Inserting Eq. (\ref{FPE particle in potentials}) into Eq. (\ref{preHJM particle in potentials}), the second line is vanishing to give 
\begin{equation}
\partial_tS+\frac{g^{ij}}{2}(\partial_{q_i}S-A_i)(\partial_{q_j}S-A_j)+V-\frac{\lambda^2}{2}\Big(g^{ij}\frac{\partial_{q_i}\partial_{q_j}R}{R}+\partial_{q_i}g^{ij}\frac{\partial_{q_j}R}{R}\Big)=0.
\label{HJM particle in potentials}
\end{equation}
We have thus pair of coupled equations (\ref{FPE particle in potentials}) and (\ref{HJM particle in potentials}) which are parameterized by $\lambda$. 

Now let us assume that $\Omega(q,\lambda;t)$ has the following symmetry: 
\begin{equation}
\Omega(q,\lambda;t)=\Omega(q,-\lambda;t),
\label{amplitude symmetry} 
\end{equation}
so that the probability density of $\lambda$ is unbiased $P(\lambda)=\int dq\Omega(q,\lambda;t)=P(-\lambda)$.
In this case, $S(q,\lambda;t)$ and $S(q,-\lambda;t)$ satisfy the same differential equation of (\ref{HJM particle in potentials}): namely the last term of Eq. (\ref{HJM particle in potentials}) is not sensitive to the signs of $\lambda$. Hence, if initially one has $S(q,\lambda;0)=S(q,-\lambda;0)$, then the symmetry will be preserved for all the time 
\begin{equation}
S(q,\lambda;t)=S(q,-\lambda;t).
\label{phase symmetry} 
\end{equation}
These properties can then be used to eliminate the last term of Eq. (\ref{FPE particle in potentials}): taking the case when $\lambda$ is positive add to it the case when $\lambda$ is negative and divided by two, one gets 
\begin{equation}
\partial_t\Omega+\partial_{q_i}\Big(g^{ij}(\partial_{q_j}S-A_j)\Omega\Big)=0. 
\label{QCE particle in potentials}
\end{equation}
We have thus pair of coupled equations (\ref{HJM particle in potentials}) and (\ref{QCE particle in potentials}) which are still parameterized by $\lambda$. 

Next, since $\lambda$ is assumed to be non-vanishing, one can define the following complex-valued function:
\begin{equation}
\Psi(q,\lambda;t)\doteq R\exp\Big(\frac{i}{|\lambda|}S\Big). 
\label{general wave function}
\end{equation}
It differs from the Madelung transformation in which $S$ is divided by $|\lambda|$ instead of $\hbar$. Equations (\ref{HJM particle in potentials}) and (\ref{QCE particle in potentials}) can thus be recast into the following modified Schr\"odinger equation: 
\begin{eqnarray}
i|\lambda|\partial_t\Psi=\frac{1}{2}(-i|\lambda|\partial_{q_i}-A_i)g^{ij}(q)(-i|\lambda|\partial_{q_j}-A_j)\Psi+V\Psi. 
\label{generalized Schroedinger equation particle in potentials}
\end{eqnarray}
Here we have assumed that the fluctuations of $\lambda$ in space and time are ignorable as compared to that of $S$. Let us proceed to assume that $\Omega$ is separable $\Omega(q,\lambda;t)=\rho(q;t)P(\lambda)$, where $P(\lambda)$ takes the form given by Eq. (\ref{God's coin}). In this case, Eq. (\ref{generalized Schroedinger equation particle in potentials}) reduces into the celebrated Schr\"odinger equation
\begin{equation}
i\hbar\partial_t\Psi_Q(q;t)=\hat{H}\Psi_Q(q;t),\hspace{2mm}\mbox{with}\hspace{2mm}\Psi_Q(q;t)=\sqrt{\rho}e^{\frac{i}{\hbar}S_Q(q;t)},\hspace{2mm}\mbox{and}\hspace{2mm} S_Q(q;t)=S(q,\pm\hbar;t), 
\label{Schroedinger equation particle in potentials} 
\end{equation}
where the quantum Hamiltonian $\hat{H}$ is given by 
\begin{equation}
\hat{H}=\frac{1}{2}(\hat{p}_i-A_i)g^{ij}(q)(\hat{p}_j-A_j)+V,\hspace{2mm}\mbox{with}\hspace{2mm}\hat{p}_i\doteq -i\hbar\partial_{q_i}. 
\label{quantum Hamiltonian particle in potentials}
\end{equation}
From Eq. (\ref{Schroedinger equation particle in potentials}) we know that the Born's statistical interpretation of wave function is valid by construction for all time, $\rho(q;t)=|\Psi_Q(q;t)|^2$. 

Let us mention that there are many approaches to derive Schr\"odinger equation with quantum Hamiltonian of the type of Eq. (\ref{quantum Hamiltonian particle in potentials}) \cite{Nelson stochastic mechanics}. The advantage of our approach is that it can be applied directly as soon as the classical Hamiltonian (or Lagrangian) is given, if a solution exists. Our method of deriving the Schr\"odinger equation can thus be regarded as to provide a method of quantization of a classical Hamiltonian. For the case $g_{ij}=m\delta_{ij}$ where $m$ is constant so that $g^{ij}=\delta_{ij}/m$, we regain the result of canonical quantization. For the case where $g^{ij}$ depends on $q$, then in contrast to canonical quantization which leads to infinite different alternatives with different ordering of operators, our model imposes a unique ordering. The same ordering is also obtained in Ref. \cite{Hall ordering}. However in contrast to the method reported there which can only be applied to classical Hamiltonian without linear term in classical momentum, our method can be applied to such a case. 

Moreover, in this specific case where $\Omega$ is separable $\Omega(q,\lambda;t)=\rho(q;t)P(\lambda)$ and $P(\lambda)$ is given by Eq. (\ref{God's coin}), Eq. (\ref{QCE particle in potentials}) becomes
\begin{equation}
\partial_t\rho+\partial_{q_i}\Big(g^{ij}(\partial_{q_j}S_Q-A_j)\rho\Big)=0. 
\label{QC CE}
\end{equation}
We can then identify an ``effective'' velocity field given by  
\begin{equation}
v_e^i=g^{ij}(\partial_{q_j}S_Q-A_j). 
\label{effective velocity}
\end{equation}
Since $S_Q(q;t)$ is just the phase of the Schr\"odinger wave function, it turns out that the above effective velocity field is equal to the ``actual'' velocity of particle (beable) in pilot-wave theory \cite{Bohm paper}. This allows us to draw a conclusion that our model will reproduce the prediction of pilot-wave theory on statistical wave-like interference pattern in slits experiment and tunneling over potential barrier \cite{interference and tunneling in PWT}. However unlike pilot-wave theory, in our model, the dynamics is strictly stochastic, the wave function is not physically real-field and the Born's statistical interpretation of wave function is valid for all time by construction. 

\section{Measurement without wave function collapse and external observer}

Now let us apply the quantization method developed in the previous Section to a class of von Neumann measurement model. To do this, let us consider the dynamics of two interacting particles, the first particle with coordinate $q_1$ represents the system whose properties being measured, and the second particle with coordinate $q_2$ represents the measuring apparatus. Let us first discuss how classical dynamics describes the whole system. Let us suppose that one wants to measure a quantity $\underline{A}_1=\underline{A}_1(q_1,{\underline{p}}_1)$. To do this, let us choose the following classical measurement-interaction Hamiltonian: 
\begin{equation}
\underline{H}=g\underline{A}_1(q_1,{\underline{p}}_1){\underline{p}}_2, 
\label{classical Hamiltonian with measurement-interaction}
\end{equation}
where $g$ is a coupling constant. Let us further assume that the interaction is impulsive so that the individual free Hamiltonians of the particles are ignorable. 

In classical mechanics we first have $d\underline{A}_1/dt=\{\underline{A}_1,\underline{H}\}=0$. $\underline{A}_1$ is thus conserved during the measurement-interaction. The idea is then to correlate the value of $\underline{A}_1(q_1,{\underline{p}}_1)$ with the classical momentum of the apparatus ${\underline{p}}_2$ while keeping the value of $\underline{A}_1(q_1,{\underline{p}}_1)$ remained unchanged. On the other hand, one also has $dq_2/dt=\{q_2,\underline{H}\}=g\underline{A}_1$, which can be integrated to give 
\begin{equation}
q_2(T)=q_2(0)+g\underline{A}_1T, 
\end{equation}
where $T$ is time span of the measurement-interaction. The value of $\underline{A}_1$ prior to the measurement can thus be inferred from the observation of the initial and final values of $q_2$. In other words, the position of the second particle plays the role of the pointer of the measurement apparatus. The above model is of course far from realistic. Especially, our model excludes the irreversibility of the registration process which can only be done by realistic apparatus plus bath with large degree of freedom. See for example Ref. \cite{Theo} for a realistic model of measurement. Here we shall focus on the issue of wave function collapse.  
  
Now let us consider an ensemble of identically prepared system whose classical Hamiltonian is given by Eq. (\ref{classical Hamiltonian with measurement-interaction}), and quantize it as in the previous section. For concreteness, without loosing generality, let us consider measurement of angular momentum. To make explicit the three dimensional nature of the problem, let us put  $q_1=(x_1,y_1,z_1)$. First let us consider the measurement of $z-$part angular momentum of the first particle 
\begin{equation}
\underline{L}_{z_1}=x_1{\underline{p}}_{y_1}-y_1{\underline{p}}_{x_1}, 
\label{classical angular momentum}
\end{equation}
where ${\underline{p}}_{x_1}$ is the conjugate momentum of $x_1$ and so on. The measurement-interaction classical Hamiltonian of Eq. (\ref{classical Hamiltonian with measurement-interaction}) thus reads
\begin{equation}
\underline{H}_l=g\underline{L}_{z_1}\underline{p}_2=g(x_1{\underline{p}}_{y_1}-y_1{\underline{p}}_{x_1}){\underline{p}}_2.
\label{classical Hamiltonian angular momentum}
\end{equation}

The Hamilton-Jacobi equation of (\ref{Hamilton-Jacobi equation}) then becomes
\begin{equation}
\partial_t\underline{S}+g\big(x_1\partial_{y_1}\underline{S}-y_1\partial_{x_1}\underline{S}\big)\partial_{q_2}\underline{S}=0. 
\label{H-J equation angular momentum}
\end{equation}
On the other hand, substituting Eq. (\ref{classical Hamiltonian angular momentum}) into Eq. (\ref{classical velocity field}), the classical velocity field is given by 
\begin{eqnarray} 
\dot{x}_1(\underline{S})=-gy_1\partial_{q_2}\underline{S},\hspace{2mm}\dot{y}_1(\underline{S})=gx_1\partial_{q_2}\underline{S},\hspace{2mm}\dot{z}_1(\underline{S})=0,\hspace{2mm}
\dot{q}_2(\underline{S})=g\big(x_1\partial_{y_1}\underline{S}-y_1\partial_{x_1}\underline{S}\big). 
\label{classical velocity angular momentum}
\end{eqnarray} 
The continuity equation of (\ref{continuity equation}) then becomes
\begin{eqnarray}
\partial_t\underline{\rho}-gy_1\partial_{x_1}(\underline{\rho}\partial_{q_2}\underline{S})+gx_1\partial_{y_1}(\underline{\rho}\partial_{q_2}\underline{S})+gx_1\partial_{q_2}(\underline{\rho}\partial_{y_1}\underline{S})-gy_1\partial_{q_2}(\underline{\rho}\partial_{x_1}\underline{S})=0. 
\label{continuity equation angular momentum}
\end{eqnarray}
Next, keeping in mind Eq. (\ref{classical velocity angular momentum}), the rule of replacement of Eq. (\ref{fundamental equation}) thus becomes
\begin{eqnarray}
\underline{\rho}\mapsto\Omega,\hspace{2mm}
\partial_{x_1}\underline{S}\mapsto\partial_{x_1}S+\frac{\lambda}{2}\frac{\partial_{x_1}\Omega}{\Omega},\hspace{2mm}
\partial_{y_1}\underline{S}\mapsto\partial_{y_1}S+\frac{\lambda}{2}\frac{\partial_{y_1}\Omega}{\Omega},\hspace{2mm}
\partial_{q_2}\underline{S}\mapsto\partial_{q_2}S+\frac{\lambda}{2}\frac{\partial_{q_2}\Omega}{\Omega},\hspace{2mm}\nonumber\\
\partial_{t}\underline{S}\mapsto\partial_{t}S+\frac{\lambda}{2}\frac{\partial_{t}\Omega}{\Omega}+ g\lambda(x_1\partial_{y_1}\partial_{q_2}S-y_1\partial_{x_1}\partial_{q_2}S).\hspace{30mm}
\label{fundamental equation angular momentum}
\end{eqnarray}

Let us proceed to see how Eq. (\ref{fundamental equation angular momentum}) modifies Eqs. (\ref{H-J equation angular momentum}) and (\ref{continuity equation angular momentum}). Imposing the first four equations of (\ref{fundamental equation angular momentum}) into Eq. (\ref{continuity equation angular momentum}) one obtains, after a simple calculation
\begin{eqnarray}
\partial_t\Omega-gy_1\partial_{x_1}(\Omega\partial_{q_2}S)+gx_1\partial_{y_1}(\Omega\partial_{q_2}S)+gx_1\partial_{q_2}(\Omega\partial_{y_1}S)\nonumber\\
-gy_1\partial_{q_2}(\Omega\partial_{x_1}S)-g\lambda(y_1\partial_{x_1}\partial_{q_2}\Omega-x_1\partial_{y_1}\partial_{q_2}\Omega)=0.
\label{FPE angular momentum}
\end{eqnarray}
On the other hand, imposing the last four equations of (\ref{fundamental equation angular momentum}) into Eq. (\ref{H-J equation angular momentum}), one has, after an arrangement
\begin{eqnarray}
\partial_tS+g\big(x_1\partial_{y_1}S-y_1\partial_{x_1}S\big)\partial_{q_2}S-g\lambda^2\Big(x_1\frac{\partial_{y_1}\partial_{q_2}R}{R}-y_1\frac{\partial_{x_1}\partial_{q_2}R}{R}\Big)\nonumber\\+\frac{\lambda}{2\Omega}\Big(\partial_t\Omega-gy_1\partial_{x_1}(\Omega\partial_{q_2}S)
+gx_1\partial_{y_1}(\Omega\partial_{q_2}S)+gx_1\partial_{q_2}(\Omega\partial_{y_1}S)-gy_1\partial_{q_2}(\Omega\partial_{x_1}S)\nonumber\\
-g\lambda(y_1\partial_{x_1}\partial_{q_2}\Omega-x_1\partial_{y_1}\partial_{q_2}\Omega)\Big)=0,
\label{ccc}
\end{eqnarray}
where we have again defined a real-valued function $R\doteq\sqrt{\Omega}$ and used the identity of Eq. (\ref{fluctuations decomposition}). Substituting Eq. (\ref{FPE angular momentum}) into Eq. (\ref{ccc}), the last term in the bracket vanishes to give 
\begin{eqnarray}
\partial_tS+g\big(x_1\partial_{y_1}S-y_1\partial_{x_1}S\big)\partial_{q_2}S-g\lambda^2\Big(x_1\frac{\partial_{y_1}\partial_{q_2}R}{R}-y_1\frac{\partial_{x_1}\partial_{q_2}R}{R}\Big)=0.
\label{HJM angular momentum}
\end{eqnarray}
The dynamics of ensemble of trajectories is then determined by pair of coupled Eqs. (\ref{FPE angular momentum}) and (\ref{HJM angular momentum}) which are parameterized by the hidden random variable $\lambda$.   

Again, assuming the symmetry of Eq. (\ref{amplitude symmetry}), one can see that $S(q,\lambda;t)$ and $S(q,-\lambda;t)$ satisfy the same differential equation of (\ref{HJM angular momentum}). Hence, assuming that initially $S(q,\lambda;0)=S(q,-\lambda;0)$, one obtains $S(q,\lambda;t)=S(q,-\lambda;t)$. As in the previous section, this can then be  used to eliminate the last term on the left hand side of Eq. (\ref{FPE angular momentum}) to give
\begin{eqnarray}
\partial_t\Omega-gy_1\partial_{x_1}(\Omega\partial_{q_2}S)+gx_1\partial_{y_1}(\Omega\partial_{q_2}S)+gx_1\partial_{q_2}(\Omega\partial_{y_1}S)-gy_1\partial_{q_2}(\Omega\partial_{x_1}S)=0.
\label{quantum CE angular momentum}
\end{eqnarray}
Further, defining complex-valued wave function as in Eq. (\ref{general wave function}), Eqs. (\ref{HJM angular momentum}) and (\ref{quantum CE angular momentum}) can then be rewritten into the following modified Schr\"odinger equation:
\begin{equation} 
i|\lambda|\partial_t\Psi=-g\lambda^2\big(x_1\partial_{y_1}-y_1\partial_{x_1}\big)\partial_{q_2}\Psi=g\frac{\lambda^2}{\hbar^2}\hat{L}_{z_1}\hat{p}_2\Psi,
\label{generalized Schroedinger equation angular momentum}
\end{equation}
where $\hat{L}_{z_1}\doteq -i\hbar(x_1\partial_{y_1}-y_1\partial_{x_1})$ and $\hat{p}_2\doteq-i\hbar\partial_{q_2}$ are the quantum mechanical $z-$angular momentum and linear momentum operators pertaining to the wave functions of the first and second particle, respectively, and again we have assumed that the spatiotemporal fluctuations of $\lambda$ is ignorable as compared to that of $S$. Finally, assuming that $\Omega(q,\lambda;t)=\rho(q;t)P(\lambda)$ is separable and taking the case when $P(\lambda)$ is given by Eq. (\ref{God's coin}), the above modified Schr\"odinger equation reduces into 
\begin{equation}
i\hbar\partial_t\Psi_Q=\hat{H}_l\Psi_Q, \hspace{2mm}\mbox{with}\hspace{2mm}  
{\hat H}_l\doteq g{\hat L}_{z_1}{\hat p}_2,
\label{quantum Hamiltonian angular momentum}
\end{equation} 
where $\Psi_Q(q;t)$ is the Schr\"odinger wave function defined as in Eq. (\ref{Schroedinger equation particle in potentials}). This result can be extended to the measurement of angular momentum along the $x-$ and $y-$ directions by cyclic permutation of $(x,y,z)$. In this case, $\hat{L}_{z_1}$ in Eq. (\ref{quantum Hamiltonian angular momentum}) is replaced by $\hat{L}_{x_1}$ and $\hat{L}_{y_1}$, the quantum mechanical angular momentum operators along the $x-$ and $y-$ directions, respectively. We have thus reproduced the results of canonical quantization as a specific case of our hidden variable model. Further, in this case, Eq. (\ref{quantum CE angular momentum}) becomes
\begin{eqnarray}
\partial_t\rho-gy_1\partial_{x_1}(\rho\partial_{q_2}S_Q)+gx_1\partial_{y_1}(\rho\partial_{q_2}S_Q)+gx_1\partial_{q_2}(\rho\partial_{y_1}S_Q)-gy_1\partial_{q_2}(\rho\partial_{x_1}S_Q)=0, 
\label{QCE angular momentum}
\end{eqnarray}
from which we can extract an effective velocity field which is equal to the actual velocity field of particles in pilot-wave theory. This then allows us to follow all the argumentation of pilot-wave theory to describe measurement without wave function collapse and external observer reproducing the prediction of quantum mechanics. 

As is discussed in Ref. \cite{AgungDQM1}, the above method can also be applied to measurement of position and momentum. Let us remark however that unlike measurement of linear and angular momentum, the measurement of position is special. Namely, we can show that in this case the pair of classical quantities $\{\underline{\rho},\underline{S}\}$ and the corresponding quantum quantities $\{\rho,S_Q\}$ satisfy the same Schr\"odinger equation with quantum Hamiltonian $\hat{H}=gq_1\hat{p}_2$. Hence, there is no quantum correction to the original classical equations. In other words, unlike measurement of linear and angular momentum, the measurement of position can reveal the value of the position of the particle prior to measurement. This implies important consequence. Recall that the results of measurement of linear and angular momentum (or measurement of any physical quantities) are inferred from the position of the second particle, namely the apparatus pointer  (in reality, any model of measurement should be reducible to position measurement). Then one might argue that one needs another particle, the third particle, as the second apparatus to probe the position of the second particle (the first apparatus). Proceeding in this way thus will lead to infinite regression: one will further need the forth particle (the third apparatus) to probe the position of the third particle (the second apparatus) and so on. In our model, however, since the quantum treatment of the position measurement is equivalent to the classical treatment revealing the position of the particle prior-measurement, then the second measurement on the position of the second particle (the first apparatus) is in principle not necessary. Namely, the results of position measurement by the second, third, forth apparatuses and so on are all equal to each other. 

\section{Possible correction to quantum mechanical prediction}

We have shown in the previous two Sections that the prediction of quantum mechanics is reproduced corresponding to a specific distribution of hidden random variable $\lambda$ given by Eq. (\ref{God's coin}). It is then imperative to see the implications if one allows $|\lambda|$ to fluctuate around $\hbar$ with small yet finite width. In this case, instead of using Eqs. (\ref{Schroedinger equation particle in potentials}) or (\ref{quantum Hamiltonian angular momentum}), one has to work with Eqs. (\ref{generalized Schroedinger equation particle in potentials}) or (\ref{generalized Schroedinger equation angular momentum}), and regards them as the natural generalization of the Schr\"odinger equation. 

Let us first discuss measurement of angular momentum in ensemble of identically prepared system so that the initial wave function of the system (first particle) $\psi(q_1)$ is given by one of the eigenfunction of the angular momentum operator $\psi(q_1)=\phi_l(q_1)$, ${\hat L}_{z_1}\phi_l=l\phi_l$, where $l$ is the eigenvalue. Let us denote the initial wave function of the apparatus (second particle) by $\varphi_0(q_2)$, assumed to be sufficiently localized. The total initial wave function of the system-apparatus is thus given by 
\begin{equation}
\Psi(q;0)=\phi_l(q_1)\varphi_0(q_2). 
\label{initial wave function angular momentum}
\end{equation}
We have thus made an idealization that the initial wave function is independent of $\lambda$. Recall that in this case, according to the standard quantum mechanics, each single measurement event will give outcome $l$ with certainty (probability one). This is one of the postulate of quantum mechanics in addition to the Schr\"odinger equation. 

Let us solve Eq. (\ref{generalized Schroedinger equation angular momentum}) with the initial condition given by Eq. (\ref{initial wave function angular momentum}). To do this, let us assume that after time-span $t$ of measurement-interaction, the wave function can be written as  $\Psi(q,\lambda;t)=\phi_l(q_1)\varphi(q_2,\lambda;t)$. Inserting this into Eq. (\ref{generalized Schroedinger equation angular momentum}) and keeping in mind that ${\hat L}_{z_1}\phi_l=l\phi_l$, one has $\partial_t\varphi+gl'\partial_{q_2}\varphi=0$, with $l'(\lambda)=\frac{|\lambda|}{\hbar}l$. It can then be directly integrated with the initial condition $\varphi(q_2,\lambda;0)=\varphi_0(q_2)$ to give $\varphi(q_2,\lambda;t)=\varphi_0(q_2-gl't)$. One thus finally has  $\Psi(q,\lambda;t)=\phi_l(q_1)\varphi_0(q_2-g|\lambda|lt/\hbar)$. Hence, in each single measurement event, the wave function of the apparatus becomes correlated to the initial state of the system and is shifted an amount of $gl'(\lambda)t$. This means that at the end of each single measurement event, the initial position of the second particle (the apparatus pointer) is shifted uniformly and randomly as 
\begin{equation}
q_2(t,\lambda)=q_2(0)+gl'(\lambda)t.
\label{apparatus pointer}
\end{equation} 

Now let us interpret the above formalism in similar way as with classical measurement. As discussed in the previous section, in the latter case, after time-span of measurement-interaction $t$, the position of the apparatus-particle is shifted as $q_2(t)=q_2(0)+g\underline{L}_{z_1}t$. From this, one infers that the result of measurement to be given by $\underline{L}_{z_1}$. Similarly, it is natural to interpret Eq. (\ref{apparatus pointer}) that the outcome of each single measurement event is given by $l'(\lambda)=|\lambda|l/\hbar$. Hence, instead of obtaining a sharp value $l$ as postulated by the standard quantum mechanics, one obtains a random value $l'(\lambda)$ which depends on the value of the hidden variable $\lambda$. One can also see that when the distribution of $\lambda$ is given by Eq. (\ref{God's coin}) so that $\lambda=\pm\hbar$, then the randomness of the outcome of single measurement disappears and one regains the prediction of quantum mechanics: $l'(\pm\hbar)=l$ with probability one. For general distribution of $\lambda$ satisfying Eq. (\ref{unbiased condition}), we have thus a random correction to the prediction of quantum mechanics: even when the initial wave function of the system is given by one of the eigenfunction of the angular momentum operator, the result of each single measurement will still be random with statistical properties determined by the distribution of $\lambda$. This observation in turn leads to a finite broadening of the spectral line purely induced by the hidden random variable. Detail elaborations of this observations is reported somewhere else \cite{AgungDQM2}. 

Next, let us notice that the generalized Schr\"odinger equation of Eqs. (\ref{generalized Schroedinger equation particle in potentials}) or (\ref{generalized Schroedinger equation angular momentum}) are still linear. Hence, given two solutions $\Psi_i(q,\lambda;t)=\sqrt{\Omega}\exp(iS/|\lambda|)$, $i=1,2$, one can construct new solution through linear superposition $\Psi_{12}=a\Psi_1+b\Psi_2$, where $a$ and $b$ are complex numbers. Calculating the probability density of the position, assuming that $\Omega_i$ is separable $\Omega_i(q,\lambda;t)=\rho_i(q;t)P(\lambda)$, the interference term then takes the following form
\begin{equation}
I_{12}(q;t)\sim\sqrt{\rho_1\rho_2}\int d\lambda P(\lambda)\exp\Big(\frac{iS_1}{|\lambda|}-\frac{iS_2}{|\lambda|}\Big),
\end{equation}
which, for general type of $P(\lambda)=P(-\lambda)$ and $S_i(q,\lambda;t)$, will give different results from the quantum mechanics. In particular if $P(\lambda)$ takes a form of symmetric log-normal function then there will be Gaussian suppression of the quantum mechanical interference even in a closed system \cite{AgungDQM3}.    

\section{Conclusion and discussion}

We have proposed a quantization method based on replacement of c-number by another c-number parameterized with hidden random variable. The results of canonical quantization is reproduced if the hidden random variable $\lambda$ can only take discrete values $\pm\hbar$ with equal probability. Of course $\lambda$ can be a function of other continuous hidden random variables. Unlike canonical quantization, the method is free from operator ordering ambiguity and has a straightforward interpretation as statistical modification of classical dynamics of ensemble of trajectories. Hence, the quantum-classical correspondence is kept physically transparent in the quantization processes. The classical limit of the Schr\"odinger equation is given by the classical dynamics of ensemble of trajectories.  
  
In particular, for all the system considered, we can identity an ``effective'' velocity field which turns out to be equal to the ``actual'' velocity of the particle in pilot-wave theory. This then allows us to conclude that the model reproduces the statistical wave-like interference pattern in slits experiment; and further we can borrow all argumentation of pilot-wave theory on measurement without wave function collapse and external classical observer. However unlike pilot-wave theory, the model is stochastic, the wave function is not physically real and the Born's statistics is valid for all time by construction. Moreover, the construction is unique given the classical Lagrangian or Hamiltonian. Finally, assuming that $|\lambda|$ fluctuates around $\hbar$ with a very small yet finite width, then the model predicts small correction to the prediction of quantum mechanics. This might lead to precision test of quantum mechanics against our hidden variable model. 

It is then imperative to ask how our model will deal with Bell's no-go theory. Since our model reproduces the prediction of quantum mechanics for specific distribution of $\lambda$, then for this case, it must violate Bell inequality which implies that it is non-local in the sense of Bell \cite{Bell unspeakable}, or there is no global Kolmogorovian space which covers all the probability spaces of the incompatible measurement in EPR-type of experiments \cite{contextuality loophole}, or both. We believe that this question can be discussed only if we know the physical origin of the the general rules of replacement postulated in Eq. (\ref{fundamental equation}). To this end, a discussion on the derivation of the rules from Hamilton-Jacobi theory with a random constraint is given some where else \cite{AgungDQM4}. 

\begin{acknowledgments} 
This research is funded by the FPR program at RIKEN. 
\end{acknowledgments}

\end{document}